\documentclass[reprint,aps,prb,superscriptaddress]{revtex4-1}

\usepackage{amsmath,amsthm,amssymb}
\usepackage{bm}
\usepackage{graphicx}

\newcommand{\mA}{{\mathsf A}}

\newcommand{\transpose}{{\!\top}}

\newcommand{\cut}{{\rm cut}}

\newcommand{\dft}{{\rm dft}}

\usepackage[normalem]{ulem}
\usepackage{color}
\definecolor{epcol}{rgb}{0,0, 0.7}
\definecolor{ascol}{rgb}{0 ,0.4, 0}
\definecolor{aocol}{rgb}{0.5 ,0, 0}

\definecolor{alertcol}{rgb}{0.7,0, 0}

\begin{document}

\title{Accelerating crystal structure prediction by machine-learning interatomic potentials with active learning}

\author{Evgeny V. Podryabinkin}
\email{E.Podryabinkin@skoltech.ru}
\affiliation{Skolkovo Institute of Science and Technology,\\
	Skolkovo Innovation Center, Building 3,	Moscow, 143026, Russia}
\author{Evgeny V. Tikhonov}
\affiliation{Skolkovo Institute of Science and Technology,\\
	Skolkovo Innovation Center, Building 3,	Moscow, 143026, Russia}
\affiliation{Sino-Russian Joint Center for Computational Materials Discovery, State Key Laboratory of Solidification Processing, School of Material Science and Engineering, Northwestern Polytechnical University, Xi’an, 710072, China}
\affiliation{International Center for Materials Discovery, School of Material Science and Engineering, Northwestern Polytechnical University, Xi’an, 710072, China}
\author{Alexander V. Shapeev}%
\affiliation{Skolkovo Institute of Science and Technology,\\
	Skolkovo Innovation Center, Building 3,	Moscow, 143026, Russia}
\author{Artem R. Oganov}%
\affiliation{Skolkovo Institute of Science and Technology,\\
	Skolkovo Innovation Center, Building 3,	Moscow, 143026, Russia}
\affiliation{Moscow Institute of Physics and Technology,\\
	9 Institutskiy per., Dolgoprudny, Moscow Region, 141701, Russia}

\date{\today}

\begin{abstract}
In this article we propose a new methodology for crystal structure prediction, which is based on the evolutionary algorithm USPEX and the machine-learning interatomic potentials actively learning on-the-fly. Our methodology allows for an automated construction of an interatomic interaction model from scratch replacing the expensive DFT with a speedup of several orders of magnitude. Predicted low-energy structures are then tested on DFT, ensuring that our machine-learning model does not introduce any prediction error. We tested our methodology on a problem of prediction of carbon allotropes, dense sodium structures and boron allotropes including those which have more than 100 atoms in the primitive cell. All the the main allotropes have been reproduced and a new 54-atom structure of boron have been found at very modest computational efforts. 

\end{abstract}
\maketitle

\section{Introduction}

Crystal structure prediction consists in searching for atomic structures with lowest thermodynamic potential \cite{CSP}. Usually theoretical methods of the prediction involve two components: an algorithm sampling the configuration space and a relaxation algorithm which finds local minima of the relevant thermodynamic potential (e.g., internal energy).
For example, the USPEX algorithm \cite{Oganov:2006,Oganov:2011,Oganov:2013} is based on evolutionary structure optimization. 
The relaxed structures are sorted according to their energies, and the lowest-energy structures are used for producing the next generation of the evolutionary algorithm.
This process of producing and relaxing structures continues generation-by-generation, until the lowest-energy structure remains unchanged for a number of generations.

The success of crystal structure prediction largely depends on the choice of the Hamiltonian. 
Density functional theory (DFT) offers a sufficient accuracy in reproducing the sophisticated interaction of real atoms in crystals, however, DFT has a high computational cost.
Indeed, the complexity of DFT calculations grows cubically with the number of atoms, and in the course of structure relaxation such calculations are repeated many times. Thus, structure relaxation typically takes more than 99.9\% of the total CPU time of the prediction when using DFT. 
Therefore, in practice, prediction of a crystal structure with DFT is usually limited to systems with a few tens of atoms. Furthermore, many types of calculations are unfordable with DFT: for example, crystal structure prediction at finite temperatures, were proper sampling of the phase space is needed for computing entropies. 

Empirical interatomic potentials---a very computationally efficient alternative to DFT---can rarely be useful for predicting new materials, because their algebraic form is limited to reproducing physical properties of a few known structures that they were specifically designed for.
A promising alternative to the empirical potentials is the so-called \emph{Machine-Learning Interatomic Potentials}.
They typically have a flexible functional form that allows for systematic improvement of their accuracy by the cost of computational efficiency.
Several approaches to developing machine-learning potentials exist: neural network-based potentials \cite{BehlerParrinello2007NN, NNP2012construction, artrith2016implementation, SmithIsayevRoitverg2017ani}, Gaussian approximation potentials \cite{Bartok2010GAP, BartokKondorCsanyi2013descriptors, GAP2014} and potentials based on linear regression, e.g., \cite{Thompson2015316} and \cite{Shapeev2016-MTP}.
In particular, neural-network potentials have been used to explore transitions between different phases of Si for a range of pressures and temperatures \cite{BehlerEtAl2008Metadynamics}. 
Machine learning methods have been successfully used for crystal structure prediction problems. For example, an idea to fit a potential while structure search was suggested in \cite{wu2014}; in \cite{jacobsen2018} the authors apply active-learning techniques to predict the surface reconstructions; in \cite{jorgensen2018} Bayesian optimization was used for the problem of prediction of molecular compounds.

\section{Machine learning interatomic potentials}

In the present paper we propose a novel approach to the application of machine-learning potentials to crystal structure prediction. It is based on the moment tensor potentials (MTPs) \cite{Shapeev2016-MTP} as the machine-learning interatomic interaction model.
Briefly, MTPs assume a partitioning of the energy into contributions of each atom:
$E = \sum_{i} {V_i}$, where $i$ goes through all atoms in the structure.
Each $V_i = V\big(\bm u_i \big)$ depends on the atomic neighborhood $\bm u_i$ defined as the collection of positions the atoms relative to the $i$-th atom within a cutoff sphere of radius $R_\cut$.
MTPs provide an analytical expression for $V_i$ as a linear combination of $m$ basis functions $B_j = B_j(\bm u_i)$ with the fitting parameters $\theta=(\theta_1,\ldots,\theta_m)$:
\begin{equation}
\label{eq:linMTP}
V_i = \sum_{j=1}^m \theta_j B_j\big(\bm u_i\big).
\end{equation}
The number of basis functions, $m$, is chosen by empirically balancing the accuracy and computational efficiency of the MTP.
The basis functions satisfy all the physical symmetries (in particular, rotation invariance and invariance toward permutation of atoms of the same type) and have explicit expressions for calculation of the forces and stresses.
The parameters $\theta$ in the simplest case are found from the requirement that the predicted energies are the DFT energies, $E\big(x^{(k)}\big) \approxeq E^\dft\big(x^{(k)}\big)$, on a set of configurations that we call the \emph{training set}.
This yields a system of linear algebraic equations on the coefficients $\theta_j$:
\begin{equation} 
\label{eq:energy_fit}
\sum_{j=1}^m \theta_j 
\underbrace{
	\left[\sum_{i=1}^N B_j\big(u_i^{(k)}\big)\right]
}_{=:b_j\left(x^{(k)}\right)} = E^\dft\big(x^{(k)}\big),
\end{equation}
which in the matrix notation we write as $\mA \theta = \vec{E}^\dft$, where
\begin{equation}
\label{eq:matrix}
\mA = \begin{pmatrix}
b_1\big(x^{(1)}\big) & \ldots & b_m\big(x^{(1)}\big) \\
\vdots & \ddots & \vdots \\
b_1\big(x^{(K)}\big) & \ldots & b_m\big(x^{(K)}\big) \\
\end{pmatrix}
.
\end{equation}
The system \eqref{eq:energy_fit} is typically overdetermined, therefore its solution is found from a least-square minimization problem as 
\begin{equation}
\label{eq:ls_solution}
\theta := (\mA^\transpose \mA)^{-1} \mA^\transpose \vec{E}^\dft
.
\end{equation}

\section{learning on the fly}

There are two main ways to train and use a machine-learning potentials. The first (classical) is to train them at the offline stage and use it for calculation of the energy, forces and stresses at the online stage. 
The biggest challenge associated with this approach is related to transferability of machine-learning potentials: they do not produce reliable predictions outside their training domain. 
This appears a fatal weakness of them in the context of crystal structure prediction: since we do not know the crystal structures to be predicted, we cannot train a potential on those structures. 

The idea of \emph{active learning} comes to rescue.
An active learning algorithm \cite{AL-MTP} can detect when an MTP attempts to extrapolate outside its training domain and include those extrapolative configurations in the training set.
To formalize the concept of extrapolation, we consider an arbitrary configuration $x^*$ and note that its predicted energy can be expressed as a linear combination of the energies of configurations from the training set:
\begin{multline}\label{eq:extrapolation}
E(x^*) = \sum_{j=1}^m \theta_j b_j(x^*) = (b^*)^\transpose \cdot \theta = 
\\
=\underbrace{(b^*)^\transpose (\mA^\transpose \mA)^{-1}\mA^\transpose}_{=: c}
\vec{E}^{\dft}
= \sum_{k=1}^K c_k E^\dft(x^{(k)}).
\end{multline}
If at least one $c_k$ in \eqref{eq:extrapolation} is larger than $1$ by its absolute value, then we consider the energy calculation as linear \emph{extrapolation} outside of the training domain; otherwise, if $|c_k| \leq 1$ for all $k$, then we say that \emph{interpolation} within the training set takes place.
In the case of interpolation the calculated energy is always bounded by the energy values from the training set (and therefore is expected to be close to the DFT values), whereas extrapolation may yield non-physically low or high values of energy.
To simplify the algorithm, we assume that the training set size, $K$, equals to the number of basis functions, $m$.
Thus, to detect extrapolation while calculating $E(x^*)$ we should additionally calculate $c_k$, which requires only one additional matrix-vector multiplication $\mA (\mA^\transpose \mA)^{-1} b^* = \mA^{-1} b^*$, provided that we store $\mA^{-1}$ in our computations.
We emphasize that no additional DFT calculations are required to detect extrapolation.

\begin{figure}[htbp]
	\centering
	\includegraphics[width=3.5in]{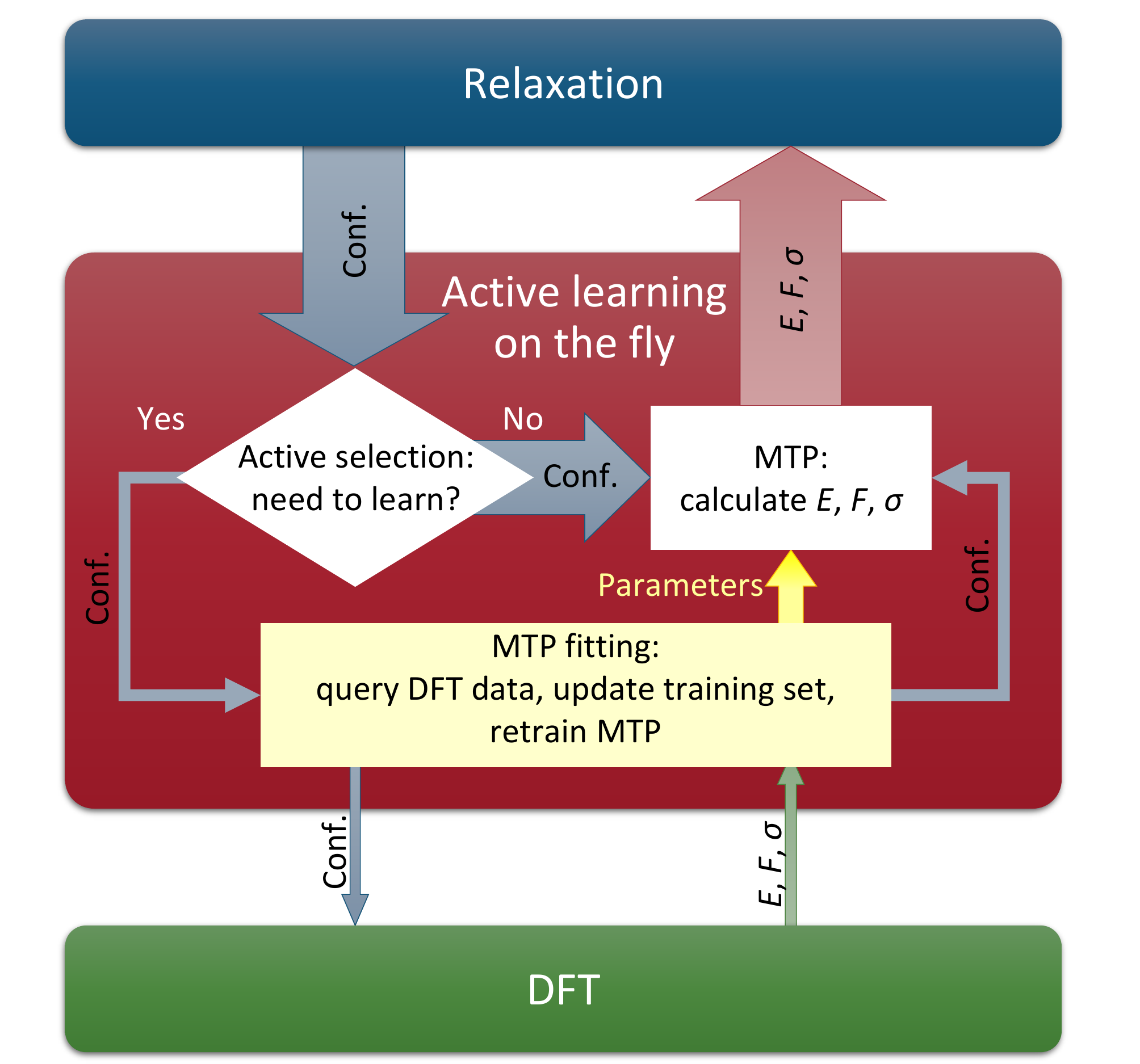}
	\caption{The scheme of learning on-the-fly.
		An active selection algorithm estimates the degree of extrapolation for each configuration sampled. If it is high then the configuration is learned.
		After this, the energy, forces and stresses are calculated by MTP and returned to the relaxation process.
	}
	\label{fig:LOTF}
\end{figure}

Our active learning approach consists of detecting and including the extrapolative configurations to the training set.
Thus, for a configuration $x^*$ we compute the extrapolation grade that we define as $\gamma(x^*) = \max_k(|c_k|)$, and compare it to the tunable parameter $\gamma_{\rm tsh} > 1$, which we call the extrapolation threshold.
If the extrapolation grade is sufficiently high, $\gamma(x^*)>\gamma_{\rm tsh}$, then the expensive DFT calculation is performed and the configuration is ``learned'', otherwise the energy, forces and stresses are calculated by the MTP (see Fig.~\ref{fig:LOTF}).
We always keep the size of the training set constant, $K=m$.
Therefore, after adding $x^*$ to the training set we eliminate $x^{(k^*)}$ from the training set, where $k^*$ is such that $c_{k^*}$ is maximal by its absolute value among all $c_k$. It can be shown that replacing configuration $x^{(k^*)}$ becomes interpolative after adding $x^*$ to the training set \cite{AL-MTP}.

The described active learning procedure employs a part of the maxvol algorithm \cite{oseledets2010how-to1770566} for finding the most linearly independent rows in a tall matrix, and it was shown that this leads to the increase of $|{\rm det}(\mA)|$ by the factor of $\gamma(x^*)>\gamma_{\rm tsh}>1$.
In other words, we select the configurations so that they maximize $|{\rm det}(\mA)|$.
Such approach is known as the D-optimality criterion \cite{settles.tr09} and is commonly used in machine learning and optimal experiment design. 

MTP that actively learns on-the-fly can be considered, effectively, as an interatomic interaction model (see Fig.~\ref{fig:LOTF}). We use this model as a replacement for DFT in USPEX. Note that since the interatomic potential within this model may slightly change when being retrained, the relaxation method requires adaptation---we force the BFGS algorithm used in relaxation \cite{Fletcher:1987:PMO:39857} to perform one gradient descent iteration once a learning event occurs. It should be emphasized that the described selection algorithm (and fitting procedure) is not limited to configurations with the same number of atoms; it allows us to actively train a single model that works on structures with different numbers of atoms.

\begin{figure}[htbp]
	\centering
	\includegraphics[width=3.4 in]{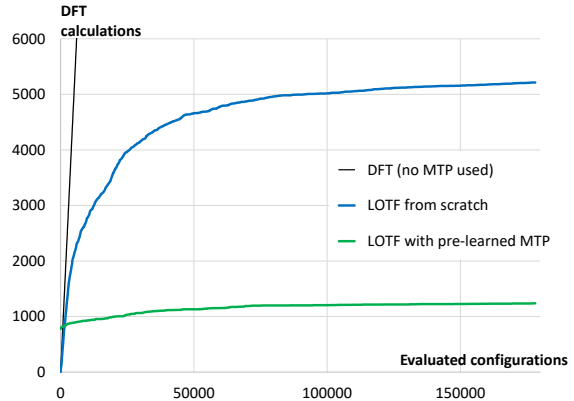}
	\caption{Comparison of learning curves for pre-trained MTP and for learning from scratch.
	}
	\label{fig:learning_curve}
\end{figure}

When an MTP starts learning on-the-fly from the empty training set, the major part of DFT calculations take place at the initial stage, while practically remain unchanged later (see the blue curve in Figure \ref{fig:learning_curve}). Such behavior is explained by the following. Our selection method unconditionally selects each configuration until the training set size is less than the number of MTP parameters $m$. Next, the configuration $x^*$ enters into the training set only if $\gamma(x^*)>\gamma_{\rm tsh}$, i.e. its extrapolation grade is larger than the threshold.
Since initially the configurational space is not well explored, the chance to meet an extrapolative configuration at the initial stage of learning on-the-fly is higher than on a later stage.
Thus, on a late stage the training set is selected from a sequence sampled from a large configurational space and hence new extrapolative configurations appear rarely.

One can take advantage of this fact by performing pre-exploration of the configurational space, thus significantly decreasing the number of DFT calculations at the initial stage.
Since our selection method does not require DFT data, the pre-exploration is done by sampling random structures subject to a minimal distance constraint.
In our tests we generated a pool of $100\,000$ structures with the different number of atoms.
The initial training thus comprises $m$ (same as the number of MTP parameters) configurations selected from the pool.
This reduces the number of DFT calculations by about a factor of 5 (compare the blue and green curve in Figure \ref{fig:learning_curve}).

\section{Crystal structure prediction with our methodology}

We have tested our methodology on finding the structures for three chemical elements: (1) carbon allotropes, (2) sodium structures under pressure, (3) boron allotropes.
In our tests no \emph{a priori} information about the low-energy structures was used. In all the three cases we employ an MTP with about 800 parameters providing a balance between accuracy and computational efficiency.
Since our model reproduces DFT only approximately, the structures predicted with DFT and our model may not coincide. Therefore, the structures with sufficiently low energies (within $100$ meV/atom of the lowest energy) found with our method were relaxed with DFT.
All DFT calculations were performed with the VASP package \cite{VASP1,VASP3,VASP4} at the generalized gradient approximation level of theory \cite{PBE} using the projected augmented wave wave method for describing effects of core electrons \cite{Blochl1994PAW, PhysRevB.59.1758}. 

\subsection{Carbon structures}
	
In the course of searching of carbon structures with 8 atoms in the unit cell our method has correctly predicted all main allotropes: graphite, diamond and lonsdaleite. More than $1.9\cdot10^4$ configurations were evaluated by MTP whereas the number of DFT calculations was about 1300.
After learning on-the-fly, the MTP training error was 86 meV/atom.
We note that the training error evaluated on the actively selected training set is an overestimation of the actual prediction error since the actively selected configurations tend to sample more ``extreme'' parts of the configurational space \cite{AL-MTP}. The error on the predicted structures was less than 40 meV/atom.

\subsection{Sodium under pressure}

Next we searched for sodium structures under pressure in the range 120--300 GPa with up to 20 atoms in the unit cell.
For this purpose we executed our method with a single MTP which correctly predicted all the known ground state sodium structures \cite{ma2009, gregoryanz2005, gregoryanz2008, lundegaard2009} (namely, cI16, tI19, hP4, provided in Supplemental Materials) within the specified pressure range.
It is remarkable and reassuring that MTP captures non-trivial physics here: hP4-Na is an electride, i.e., can be described as an ionic salt made of Na$^+$ cores and localized electron pairs, which play the role of anions. We found it remarkable that MTP, based only on nuclear positions, is able to recognize and predict this electride phase.
About 1500 DFT calculations have been done while learning on-the-fly, whereas the total number of evaluated configurations was more than $2\cdot10^6$.
In this test the speedup of our method (which is equal to the ratio between the number of DFT and MTP calculations) is 100 times higher than in the previous, which indicates a higher efficiency of our method when searching structures with more atoms.

\subsection{Boron allotropes}

In the third test we demonstrate an application of our approach to one of the most challenging, in our opinion, crystal structure prediction problems: finding the allotropes of boron. Indeed, boron has an extremely complex potential energy surface with large amount of local mininima (i.e., large number of allotropes) and small energy differences between them. Moreover, some of allotropes have more than 100 atoms in their primitive cell \cite{B106, oganov2009ionic, PhysRevLett.117.085501, PhysRevB.67.174116, VanSetten}, therefore, exhaustive search is virtually impossible since the number of possible structures grows exponentially with the number of atoms.

We started by looking for boron allotropes with 1--60 atoms in the primitive cell.
We started from structures with one atom in the primitive cell, and increased the number of atoms by one in each subsequent run. 
At the 12th stage of the search the $\alpha$-boron 12-atom structure was found with MTP.

\begin{figure}[htbp]
	\centering
	\includegraphics[width=2.0in]{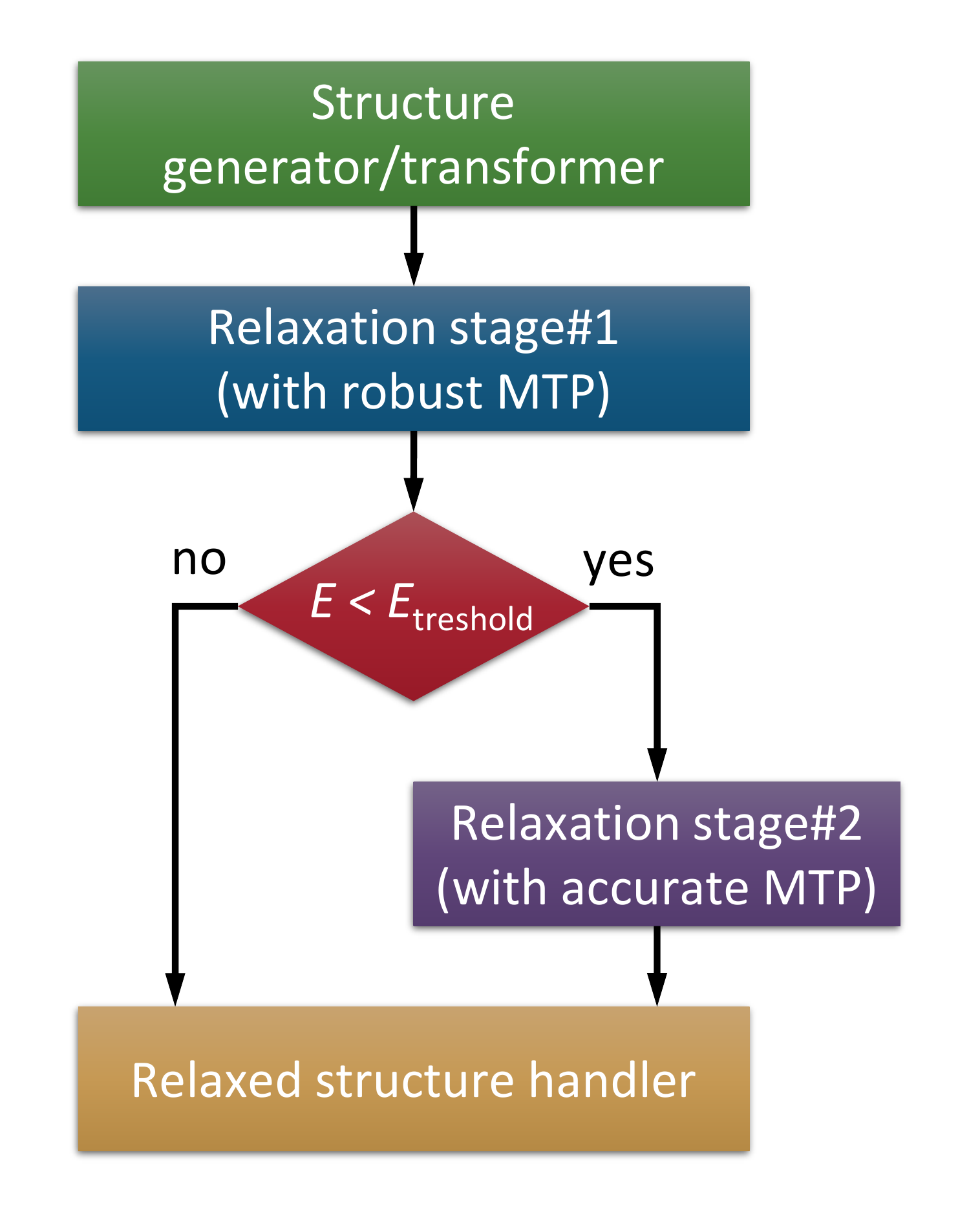}
	\caption{The two-stage relaxation scheme}
	\label{fig:2stage}
\end{figure}

Despite the fact that the $\alpha$-boron structure was correctly identified as the ground state, the training error was 170 meV/atom. This error seems large, therefore, we modified our procedure by adding the second actively learning MTP. The first, reliable, MTP is used to screen out the high-energy structures, whereas the second one was introduced for accurate treatment of configurations with sufficiently low energy per atom.
Switching between the reliable and accurate MTPs is done in a two-stage relaxation (see Fig. \ref{fig:2stage}).
At the first stage a configuration is relaxed with a reliable MTP and is discarded if its energy is higher than $-5.5$ eV/atom.
Otherwise, the relaxation proceeds with the accurate MTP.
Active learning of the accurate MTP is further restricted to low-energy structures by applying to the selection and training procedures the weight 
\[W(x) = 1 / (E(x)/N - E_{\rm min} + 0.02)^2,\]
where $E_{\rm min}$ eV/atom is the 
minimal DFT per atom energy of the structures from the training set ($E_{\rm min}=-6.705$ after finding of $\alpha$-boron), and $N$ is the number of atoms in the configuration $x$.
That is, both sides of every equation \eqref{eq:energy_fit} are multiplied by $W\big(x^{(k)}\big)$.
Being trained only on low-energy configurations the second MTP better reproduces the potential energy surface near deep minima, but is not suitable for high-energy structures.

The two-stage scheme yields the root-mean-square error of 11 meV/atom for the low-energy structures (we have compared the MTP and DFT energies for the 100 lowest-energy structures we found).
The total number of DFT calculations required to train the two MTPs was about 5000  while the total number of evaluated configurations exceeds $4\cdot10^8$. The best structures (with the lowest energy) found with our method are shown in Figure~\ref{fig:structs}.
The data files for these structures 
are provided in Supplemental Materials. 

\begin{figure}[htbp]
	\centering
	\includegraphics[width=3.6 in]{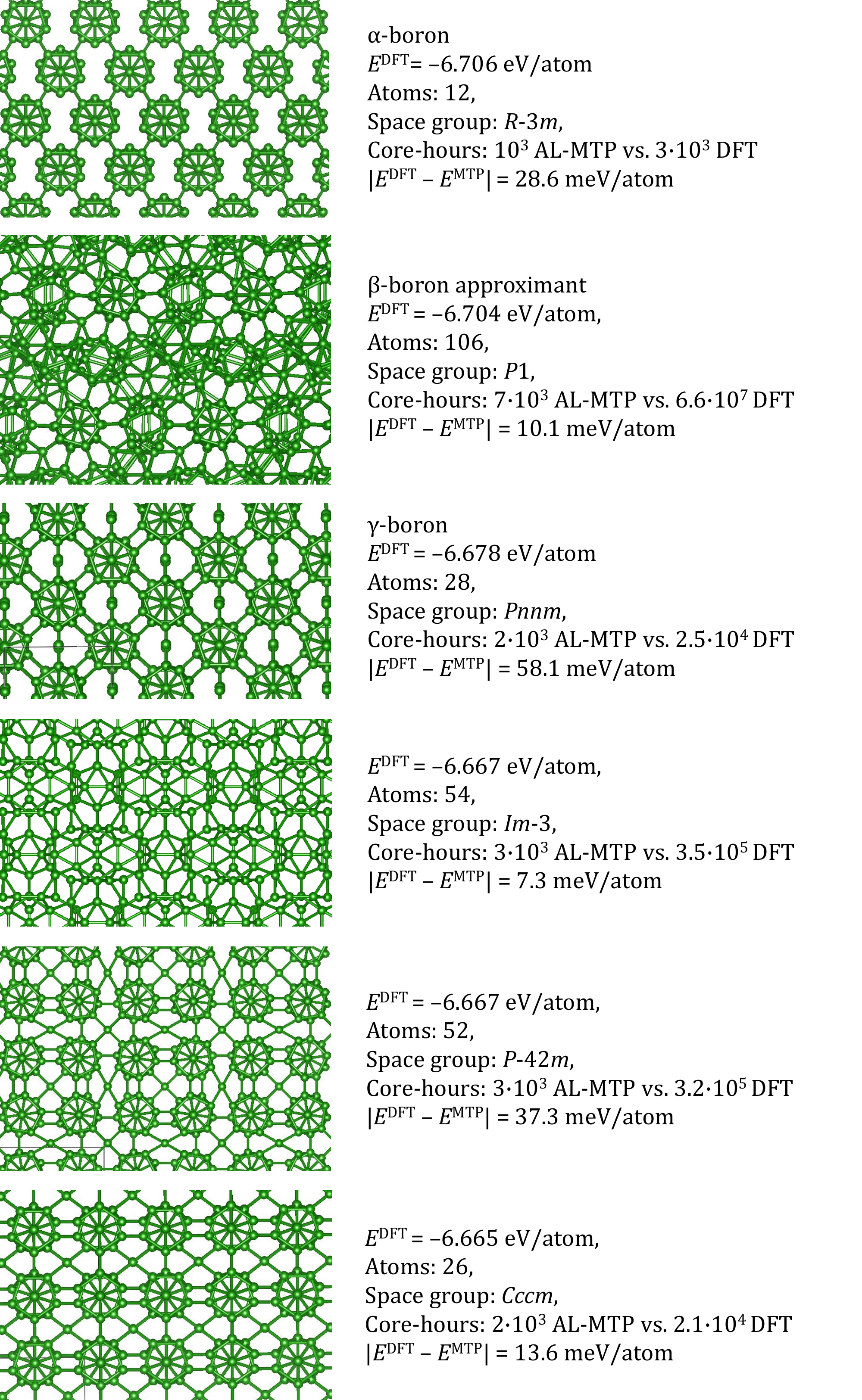}
	\caption{The best (lowest-energy) structures found with our method. The estimate for the time required to find these structures on DFT is based on the number of configuration threated by MTP in our search and the time required for VASP to process all these structures on a single core.
		The actual time spent with DFT can be up to 10 times less than indicated, because at early stages of structure relaxation cheaper computational settings are usually applied.
	}
	\label{fig:structs}
\end{figure}

The USPEX+MTP calculation correctly reproduced the lowest-energy $Cccm$ structure with 26 atoms/cell and also discovered a closely related tetragonal $P{\bar{4}2m}$ structure with 52 atoms/cell and slightly lower energy. These two structures are topologically very similar to the 52-atom $Pnn2$ structure published in \cite{Zhu:sn5112} and the 52-atom $Pnnn$ structure recently seen in experiments \cite{ekimov2016}, the fourth established pure boron allotrope. These structures are very similar to the tetragonal ($P4_2/nnm$) B$_{50}$, the first structurally characterized form of boron \cite{Hoard1951}, which was later found to be impure and stabilized by nitrogen and carbon atoms \cite{Ploog1971}. 
DFT predicts that all the three 52-atom structures and the 26-atom structure are dynamically stable at zero temperature and have very close energies, within 8 meV/atom (all four structures are provided in Supplemental Materials).
However, MTP ``catches'' only two of them: the $Cccm$ and $P$-42$m$ structures. Relaxation of $Pnn2$ and $Pnnn$ structures results in the 26$\times$2-atom $Cccm$ structure.
This proves boron to be a hard benchmark problem for crystal structure prediction.
With our two-MTP scheme we have also predicted correctly the $\gamma$-boron structure, recently discovered theoretically and experimentally \cite{oganov2009ionic}. 

A remarkable result of this work is the discovery of a 54-atom metallic structure of boron, with energy just 39 meV/atom higher than that of $\alpha$-boron and 11 meV/atom higher than that of $\gamma$-boron. This structure has an unexpectedly high symmetry, cubic (space group $Im$-3), the point group of which is actually the highest group that an icosahedron can have in any crystalline environment ($m$-3). The B$_{12}$ icosahedron is connected to the neighboring icosahedra only by two-center bonds---this is what allows it to keep the highest symmetry allowed for icosahedron in crystals.
As a result, we can identify in this structure the highly symmetric B$_{84}$ units: in their center is one B$_{12}$ icosahedron, with each of its atoms bonded by a two-center bond to a B$_6$ pentagonal “umbrella”. These B$_6$-umbrellas can be viewed as fragments of icosahedra; building this structure with complete icosahedra would lead to an aperiodic and highly strained structure; our B$_{54}$ is its simplest periodic approximant.
 
We have checked thermal stability of this structure with molecular dynamics simulations. To this end we replicated the unit cell and ran simulations with NPT ensemble of 13500 atoms at $T = 1200 K$ with the accurate MTP. After 100 ps we performed relaxation and verified that no structure transformation occurred. The structure has specific volume of 7.61 \AA$^3$/atom, which is close to that of $\beta$-boron. It has the following elastic constants: $C_{11}=418$ GPa, $C_{12}=102$ GPa, $C_{44}=160$ GPa, bulk modulus $B=208$ GPa, shear modulus $G=160$ GPa and Young's modulus $E=381$ GPa. It is a brittle metallic structure according to Pugh's criterion \cite{PUGH} (with $B/G$ ratio of $1.3$) with the hardness of $H=25.3$ GPa calculated with the formula proposed in \cite{CHEN20111275}. Low mass of the boron and covalent B-B bonding (leading to strong electron-phonon coupling) make metallic boron a potential superconductor, however, here we found $T_c<1 K$.

\subsection{Prediction of $\beta$-boron structure}

The most challenging test for our methodology is an attempt to predict the $\beta-$boron structure. It has a high configurational entropy and even now its exact structure remain unclear \cite{PhysRevLett.117.085501}. The minimal number of atoms in the primitive cell of such approximants is in the range of 105--108.     
Up to now, all theoretical attempts to predict the structure of $\beta-$boron \cite{widom2008, VanSetten, B106, gabor2018} used experimental knowledge of lattice parameters and of positions of most boron atoms (the fully occupied sites), only varying the occupation of the partially occupied sites. Here we present the first attempt to perform a fully theoretical and unconstrained prediction of this structure, using no empirical information (except for the approximate number of atoms in the supercell). The success that we have achieved is remarkable and sets the record of complexity of the predicted crystal structure.

\begin{figure}[htbp]
	\centering
	\includegraphics[width=2.0 in]{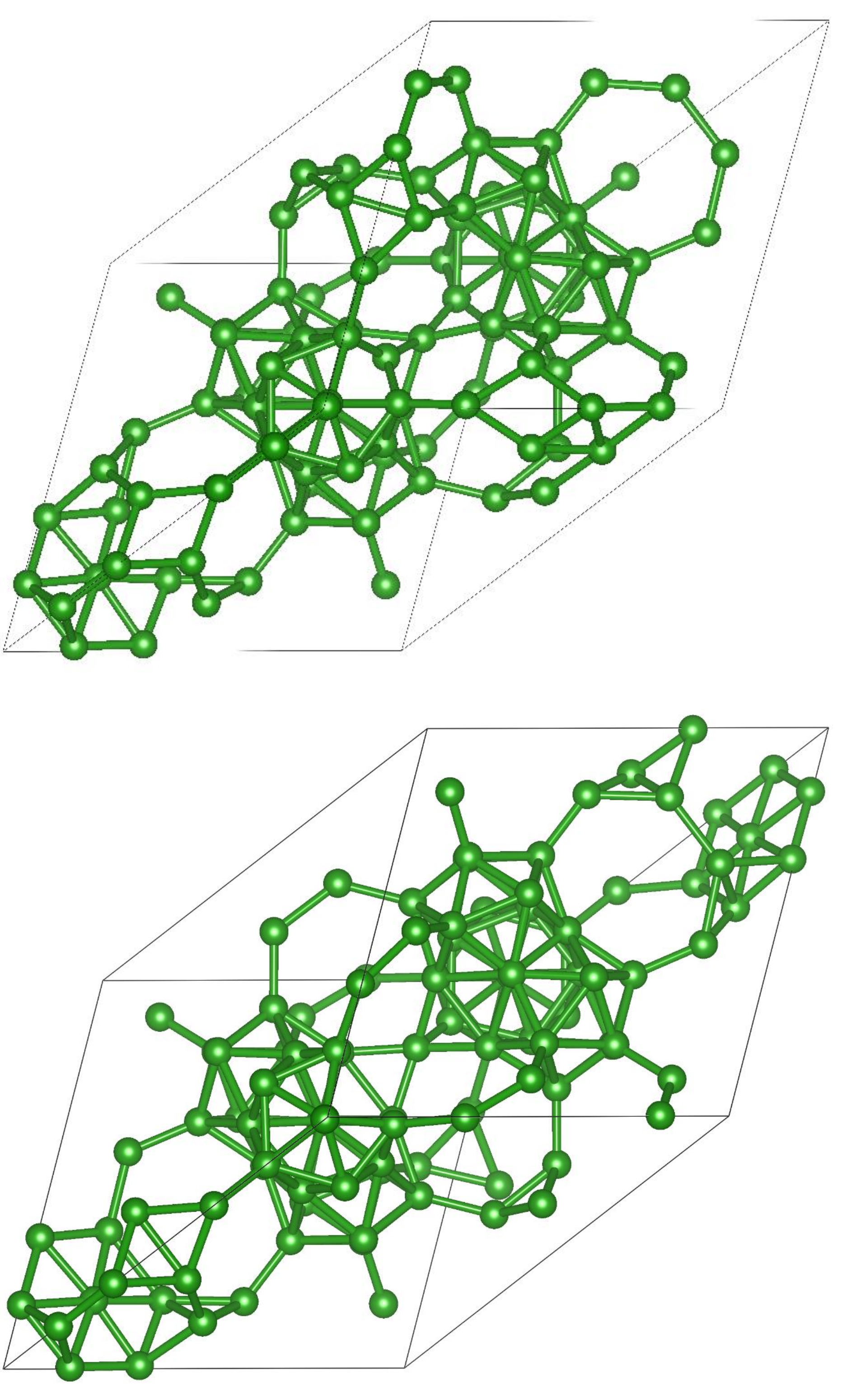}
	\caption{Comparison of found 106-atom structure (upper figure) and structure from \cite{VanSetten} (lower figure).
	}
	\label{fig:B106}
\end{figure}

With our method we have found the best known 108-atom structure of boron, which is a supercell of $\alpha$-boron (the same allotrope was also found among all the appropriate smaller structures). It confirms that our method is able to treat large structures with more than 100 atoms in the primitive cell.
Furthermore, another predicted structure with 108 atoms in the unit cell has structural similarity with $\beta$-boron \cite{VanSetten} (in particular, it also contains fused icosahedra with point defects and some non-icosahedral atoms, refer to Supplemental Materials) and energy only 8 meV/atom higher than that of $\alpha$-boron.
In addition to the 108-atom $\alpha$-boron supercell structure, we have also found $\beta$-boron approximants with 105, 106, 107, and 108 atoms in the primitive cell whose energies are $14$, $2$, $8$, and $8$ meV/atom higher as compared to $\alpha$-boron.
These configurations have structural similarity with the known 106 atomic $\beta$-boron approximant \cite{VanSetten}, in particular, they also contain fused icosahedra with point defects and some non-icosahedral atoms.
Moreover, our 106-atom structure has virtually the same energy as provided in \cite{VanSetten} (DFT energies differ by less than 1 meV/atom). Both structures are shown in Fig.~\ref{fig:B106}. Interestingly, our calculations found more than hundred $\beta$-boron-like structures with close energies, which is a manifestation of its configurational entropy.

\section{Conclusion}

In summary, we have proposed and tested a methodology for crystal structure prediction, based on a machine-learning interatomic interaction model and the evolutionary algorithm USPEX.
The new methodology is orders of magnitude more computationally efficient compared to the conventional DFT-based algorithms.
Our machine-learning model is automatically trained on-the-fly and does not require manually assembling the training dataset, thus seamlessly replacing DFT without significant changes to the crystal structure prediction algorithm.
We have applied this method to sodium under pressure, and to carbon and boron. For compresed sodium, all high-pressure phases (including host-guest tI19 and electride hP4 phases) were found. For carbon, graphite, diamond, and lonsdaleite were reproduced. For boron we demonstrated that accuracy of 11 meV/atom is achievable. All known pure boron allotropes were found (including disordered $\beta-$boron with 106 atoms/cell), and a new low-energy metallic allotrope predicted.

\begin{acknowledgments}

A.R.O.\ and E.V.T.\ thanks Russian Science Foundation (grant 16-13-10459) for financial support. A.V.S.\ and E.V.P.\ acknowledge funding from the 
Russian Science Foundation (grant number 18-13-00479).
This work was performed, in part, at the Center for Integrated Nanotechnologies, an Office of Science User Facility operated for the U.S. Department of Energy (DOE) Office of Science by Los Alamos National Laboratory (Contract DE-AC52-06NA25396) and Sandia National Laboratories (Contract DE-NA-0003525).

\end{acknowledgments}

\bibliography{paper}

\end{document}